\documentclass[acmsmall,authorversion,nonacm]{acmart}
\pdfoutput=1 

\setcopyright{acmcopyright}
\copyrightyear{20XX}
\acmYear{20XX}
\acmDOI{XXXXXXX.XXXXXXX}

\acmJournal{TOMS}
\acmVolume{0}
\acmNumber{0}
\acmArticle{0}
\acmMonth{0}

\usepackage[utf8]{inputenc}
\usepackage[T1]{fontenc}

\usepackage{amsmath,amsfonts,physics}
\usepackage{float}

\usepackage{siunitx} 
\DeclareSIUnit\byte{B} 
\usepackage{tikz} 
\usetikzlibrary{calc} 
\usepackage{pgfplots} 
\pgfplotsset{compat=1.11} 
\usepackage{xurl} 
\usepackage{hyperref}
\hypersetup{breaklinks=true}
\usepackage{cleveref} 

\usepackage{listings}
\lstdefinelanguage{myasm}{
  morecomment=[l]{\#},
  morecomment=[l]{;},
  morecomment=[s]{/*}{*/},
  morestring=[b]",
  sensitive=false,
  morekeywords={movq,leaq,syscall,lock,cmpxchgq,vmovq,vaddsd,jmp,endbr64,vxorpd,vcomisd,ja,vaddsd,ret,vmovdqu16,vpbroadcastw,vpsubw,vporq,vzeroupper,movl,vmovss,fldt,fmulp,fstpt,xorq,addq,vmulsd,vmovsd,vmovupd,vmulpd,call,subq,jne,andpd,xorpd,movapd,cmplted,movsd,andpd,andnpd,orpd,addsd,retq,shrq,andl,subl,movabsq,andq,sall,vmovd,addl,orq,cmpltsd,xorl,rolq,xchgq},
  alsoletter={.:\%},
  morekeywords={.quad,.text,.data,.long,.section,.value,.globl},
  morekeywords=[2]{\%rip,\%rax,\%rdi,\%rsp,\%st,\%xmm0,\%xmm1,\%xmm2,\%zmm0,\%rdx,\%eax,\%zmm1,\%rcx,\%edi,\%edx,\%rbx},
}

\lstdefinelanguage{vex}{
    keywords = {}
}
\lstdefinelanguage{ieee754}{
    keywords = {}
}
\lstdefinelanguage{asm}{ 
    keywords = {}
}

\usepackage[skins]{tcolorbox} 

\newcommand{\vexp}[1]{{\color{red}#1}}
\newtcbox{\vexe}{%
  nobeforeafter,size=fbox,sharp corners,
  shrink tight,
  extrude by=1pt,
  tcbox raise base,
  colback=white,
  colframe=blue,
  borderline={0.9pt}{-1pt}{blue,opacity=0},
  opacityframe=0.75,
  opacityback=1,
}

\floatstyle{plaintop}
\newfloat{lstfloat}{tbp}{lop}
\floatname{lstfloat}{Listing}

\crefname{lstfloat}{Listing}{Listings}
\Crefname{lstfloat}{Listing}{Listings}

\crefname{section}{Section}{Sections}
\crefname{figure}{Figure}{Figures}
\crefname{table}{Table}{Tables}

\urldef\urldefNumPyIfThenElse\url{https://github.com/numpy/numpy/blob/8f4b73a0d04f7bebb06a154b43e5ef5b5980052f/numpy/core/src/umath/simd.inc.src#L2540}
\urldef\urldefNumPyRange\url{https://github.com/numpy/numpy/blob/8f4b73a0d04f7bebb06a154b43e5ef5b5980052f/numpy/core/src/umath/simd.inc.src#L1558}
\urldef\urldefGlibcRange\url{https://sourceware.org/git/?p=glibc.git;a=blob;f=sysdeps/ieee754/dbl-64/s_sin.c;h=8e65f7cc00faf64f4ca85a2a1e937280bcd06d3a;hb=HEAD#l156}
\urldef\urldefGlibcRoundingForTableLookup\url{https://sourceware.org/git/?p=glibc.git;a=blob;f=sysdeps/ieee754/dbl-64/s_sin.c;h=8e65f7cc00faf64f4ca85a2a1e937280bcd06d3a;hb=HEAD#l136}
\urldef\urldefGlibcTaylorEleven\url{https://sourceware.org/git/?p=glibc.git;a=blob;f=sysdeps/ieee754/dbl-64/s_sin.c;h=8e65f7cc00faf64f4ca85a2a1e937280bcd06d3a;hb=HEAD#l129}
\urldef\urldefGlibcTaylorOne\url{https://sourceware.org/git/?p=glibc.git;a=blob;f=sysdeps/ieee754/dbl-64/s_sin.c;h=8e65f7cc00faf64f4ca85a2a1e937280bcd06d3a;hb=HEAD#l215}
\urldef\urldefValgrindLibvexIR\url{https://sourceware.org/git/?p=valgrind.git;a=blob;f=VEX/pub/libvex_ir.h;h=85805bb69b8b447d0d5cd362b1fd3e5b9b181c28;hb=8b2cf214afb2590ecef5ff7cabeb6ceec3862ade}
\urldef\urldefCPython\url{https://github.com/python/cpython}
\urldef\urldefCPythonPyFloatObject\url{https://github.com/python/cpython/blob/787498cbbb7d1c7115a7af4435efb7f607b10ed1/Include/cpython/floatobject.h#L7}
\urldef\urldefCPythonOperations\url{https://github.com/python/cpython/blob/787498cbbb7d1c7115a7af4435efb7f607b10ed1/Objects/floatobject.c#L597}
\urldef\urldefShadowMemory\url{https://github.com/cronburg/shadow-memory}
\urldef\urldefDerivgrindScicomp\url{https://www.scicomp.uni-kl.de/software/derivgrind/}
\urldef\urldefDerivgrindGithub\url{https://github.com/SciCompKL/derivgrind}
\urldef\urldefShadowmemGithub\url{https://github.com/SciCompKL/derivgrind}

\newenvironment{customlegend}[1][]{%
    \begingroup
    \csname pgfplots@init@cleared@structures\endcsname
    \pgfplotsset{#1}%
}{%
    \csname pgfplots@createlegend\endcsname
    \endgroup
}%
\def\addlegendimage{\csname pgfplots@addlegendimage\endcsname}

\newcommand{\addlegendimageintext}[1]{%
    \tikz {
        \begin{customlegend}[
            legend entries={\empty},
            legend style={
                draw=none,
                inner sep=0pt,
                column sep=0pt,
                nodes={inner sep=0pt}}]
        \addlegendimage{#1}
        \end{customlegend}
    }%
}

\begin{document}

\title{Forward-Mode Automatic Differentiation of Compiled Programs}

\author{Max Aehle}
\email{max.aehle@scicomp.uni-kl.de}
\orcid{0000-0002-6739-5890}
\author{Johannes Bl{\"u}hdorn}
\email{johannes.bluehdorn@scicomp.uni-kl.de}
\orcid{0000-0002-3840-6941}
\author{Max Sagebaum}
\email{max.sagebaum@scicomp.uni-kl.de}
\orcid{0000-0001-9038-3428}
\author{Nicolas R. Gauger}
\email{nicolas.gauger@scicomp.uni-kl.de}
\orcid{0000-0002-5863-7384}
\affiliation{%
  \institution{University of Kaiserslautern-Landau (RPTU)}
  \city{Kaiserslautern}
  \country{Germany}
}


\begin{abstract}
Algorithmic differentiation (AD) is a set of techniques that provide partial derivatives of computer-implemented functions. Such a function can be supplied to state-of-the-art AD tools via its \emph{source code}, or via an intermediate representation produced while compiling its source code.

We present the novel AD tool Derivgrind, which augments the \emph{machine code} of compiled programs with forward-mode AD logic. Derivgrind leverages the Valgrind instrumentation framework for a structured access to the machine code, and a shadow memory tool to store dot values. Access to the source code is required at most for the files in which input and output variables are defined.

Derivgrind's versatility comes at the price of scaling the run-time by a factor between 30 and 75, measured on a benchmark based on a numerical solver for a partial differential equation. Results of our extensive regression test suite indicate that Derivgrind produces correct results on GCC- and Clang-compiled programs, including a Python interpreter, with a small number of exceptions. While we provide a list of scenarios that Derivgrind does not handle correctly, nearly all of them are academic counterexamples or originate from highly optimized math libraries. 
As long as differentiating those is avoided, Derivgrind can be applied to an unprecedentedly wide range of cross-language or partially closed-source software with little integration efforts.
\end{abstract}



\received{XXX}
\received[revised]{XXX}
\received[accepted]{XXX}

\maketitle


\newcommand{\Cpp}{C{\ttfamily ++}}

\newcommand{\DerivGrind}{Derivgrind}

\newcommand{\RR}{{\mathbb R}}

\section{Introduction}\label{sec:introduction}
In many areas of science and technology, processes of interest can be described by a function $f: \RR^n \to \RR^m, x\mapsto y$  implemented in a computer program. In order to leverage such a simulation for, e.\,g., gradient-based optimization,  knowledge of the derivatives $\tfrac{\partial y_i}{\partial x_j}$ is required.
To this end, besides numerical and symbolic diffentiation, algorithmic differentiation (AD) \cite{griewank_evaluating_2008} has found widespread use in various areas such as aerodynamic shape optimization \cite{luers_adjoint-based_2018} and machine learning \cite{baydin_automatic_2017}. AD is a set of techniques to turn a computer program for $f$ into a floating-point accurate computer program for $\tfrac{\partial y_i}{\partial x_j}$. Both the \emph{forward} and \emph{reverse} mode of AD run the original program $f$, and perform additional AD logic before, alongside, and afterwards. This work is concerned with the forward mode in the case of a single ($n=1$) input variable $x$: For each input, intermediate, and output value $a$, the AD logic additionally computes its \emph{dot} or \emph{tangent} value $\dot a = \tfrac{\partial a}{\partial x}$. The dot value $\dot x$ of the single AD input is initialized with $1$, $\dot a$ is initialized with $0$ for every other global or constant variable $a$,  and each real arithmetic operation $a_\text{lhs} = \phi(a_1, \dots, a_k)$ performed by the original program is matched by an update \begin{equation}\label{eq:forward-update}
\dot a_\text{lhs} = \frac{\partial \phi}{\partial a_1} \cdot \dot a_1 + \dots +  \frac{\partial \phi}{\partial a_k} \cdot \dot a_k.
\end{equation}
For instance, a multiplication $a_\text{lhs} = a_1 \cdot a_2$ leads to an update $\dot a_\text{lhs} = a_2 \cdot \dot a_1 + a_1 \cdot \dot a_2$ of the dot value of the product, and $a_\text{lhs} = \sin(a_1)$ should be accompanied by $\dot a_\text{lhs} = \cos(a_1) \cdot \dot a_1$. Several ways for the augmentation of the program with AD logic such as \eqref{eq:forward-update} have been described in the literature.
\begin{itemize}
\item \emph{Source transformation} tools like TAPENADE \cite{hascoet_tapenade_2013} parse the source code of the program; forward-mode AD logic like \eqref{eq:forward-update} can then be inserted directly into the source code.
\item \emph{Operator overloading} tools like ADOL-C \cite{Walther2012Gsw}, CoDiPack \cite{SaAlGauTOMS2019} and autograd \cite{maclaurin2015autograd} rely on a language feature of many object-oriented languages like \Cpp{} and Python. They define a special type to be used instead of the built-in floating-point types such as \lstinline[language=C,keywordstyle=\ttfamily]|double| and \lstinline[language=python]|np.array|. The new type stores both $a$ and additional AD data such as $\dot a$, and overloads all elementary operations and math functions with their original action on $a$ accompanied by additional AD logic. 
\item AD logic can also be  added during compilation. The AD tool CLAD \cite{Vassilev_Clad} is a plugin for the Clang front-end of the LLVM compiler infrastructure, operating on Clang's internal representation of the C/\Cpp{} source code in the form of an abstract syntax tree (AST). 
Enzyme \cite{NEURIPS2020_9332c513} operates on the LLVM intermediate representation (IR).
\end{itemize}

Except for Enzyme, a common weakness of these approaches is that they rely on the entire source code implementing the real arithmetic to be differentiated. Thus, they cannot be applied if these sources are not fully available, or if they are written in an exotic programming language, or a mix of languages for which no AD tool exists yet. 

Different languages are not a problem for Enzyme as long as an LLVM front-end exists for all of them. Also, Enzyme makes it possible to include functions from static libraries into AD without need for their source code, if these static libraries were shipped with embedded LLVM IR.

In this work, we go even further away from the source code. Our novel tool \emph{\DerivGrind{}}\footnote{\urldefDerivgrindScicomp{} and \urldefDerivgrindGithub} propagates dot values through  a compiled ``client'' program, dynamically augmenting portions of its machine code with forward-mode AD logic just before they execute on the processor. Users of Derivgrind only need access to a (usually small) number of source files in which they want to define AD input and output variables. \DerivGrind{} therefore enables the integration of forward-mode AD into complex, cross-language software projects involving closed-source dependencies, with minimal effort.
We target clients running under Linux on both the \emph{x86} (also referred to as i386 or IA-32) and \emph{amd64} (x86-64, Intel~64 or x64) instruction set architectures.

\DerivGrind{} has been implemented as a \emph{tool} within the Valgrind framework  for building dynamic analysis tools \cite{valgrind-paper,valgrind-doc}. After invoking
\begin{equation*}
\text{\lstinline[language=bash,mathescape]|valgrind --tool=$\langle\text{\rmfamily\emph{tool}}\rangle$ $\phantom{a}\langle\text{\rmfamily\emph{tool options}}\rangle$ $\phantom{a}\langle\text{\rmfamily\emph{client executable}}\rangle$ $\phantom{a}\langle\text{\rmfamily\emph{arguments for client}}\rangle$|}
\end{equation*}
the Valgrind core starts to read portions of the machine code of the \emph{client program}, which consists of the client executable as well as dynamically linked or loaded libraries. Valgrind translates these portions into the unified object-oriented \emph{VEX} IR, which it presents to the selected Valgrind \emph{tool}. The tool can modify (``instrument'') the VEX IR. Afterwards, the Valgrind core translates the instrumented VEX code back to machine code, which can then be executed on the CPU; we may think of this as running the instrumented VEX code on a ``synthetic CPU''.

As an example, Valgrind's default tool \emph{Memcheck} \cite{memcheck-paper} adds instrumentation to keep track of so-called availability and validity bits and detect memory errors based on this information. A standard Valgrind distribution includes several other tools for thread error detection and profiling.
Additional tools have been created independently, e.\,g.\ for taint analysis \cite{taintcheck,flayer}, crash consistency checking of persistent memory applications \cite{pmemcheck,pmat}, or the analysis of floating-point accuracy problems \cite{fpdebug,herbgrind,verrou-1,verrou-2,verrou-3}.
Our novel tool \DerivGrind{} augments the VEX IR with forward-mode AD logic.

Besides the instrumentation, Valgrind tools can specify \emph{monitor commands} and \emph{client requests} to enable communication between the tool, the user, and the client program. \DerivGrind{} uses this feature for setting and getting dot values. As Valgrind tools can wrap library functions, \DerivGrind{} provides analytical derivatives for functions from the C math library.  

This paper is structured as follows. In \cref{sec:ad-forward-theory}, we recall how compiled programs usually handle floating-point data, and how this handling can be extended to propagate dot values alongside.  \Cref{sec:instrumentation-machine-code} starts with a review of VEX IR, and describes the specific AD instrumentation added by \DerivGrind{}. \Cref{sec:interaction,sec:libm-extension} detail the monitor commands and client requests, and the function wrappers introduced by \DerivGrind{}.  We apply \DerivGrind{} in multiple case studies for validation and performance analysis in \cref{sec:application}, and close with a summary and outlook in \cref{sec:summary}.

\section{Instrumentation of Machine Code with Forward AD Logic}\label{sec:ad-forward-theory}
\subsection{Shadowing Approach}\label{sec:shadowing}
Our approach to store dot values is to match every storage location, such as memory and registers, by a \emph{shadow} storage location of the same size and type. Whenever a location stores a representation of a floating-point value $a$ in any format, the AD instrumentation should make sure that the corresponding shadow location stores the dot value $\dot a$ in the same format. This does also apply to individual bytes of a floating-point representation in case the representation is split up into several parts, e.\,g.\ to copy them separately. When the data at a location is not part of a binary representation of a floating-point value, the  content of the corresponding shadow location is unspecified.

\subsection{Floating-Point Formats and Instructions}
Among other representations of real numbers as digital data, the IEEE-754 standard \cite{ieee754}, revised in~2008, specifies the most commonly used  binary floating-point interchange formats \lstinline[language=ieee754]{binary32} (formerly: single) and \lstinline[language=ieee754]{binary64} (formerly: double). 
In the 8-byte format \lstinline[language=ieee754]{binary64},
\begin{itemize}
\item the most significant bit   stores the sign, \lstinline[language={}]|0| indicating a positive and \lstinline[language={}]|1| indicating a negative number, 
\item the~11 next-most significant bits   store the integer exponent in a biased fashion, meaning that \lstinline[language={},mathescape]|0b00$\ldots$01|  and \lstinline[language={},mathescape]|0b11$\ldots$10| represent the lowest and highest possible exponents $-1022$ and $1023$, respectively, and
\item the remaining~52 lower-significant bits  store the significand (also called mantissa), apart from its implicit leading digit~1,
\end{itemize}
so \lstinline[language={},mathescape]|0b$\beta_{63} \beta_{62}  \ldots \beta_1 \beta_0$| represents the real number
\begin{gather}
(-1)^{\beta_{63}} \cdot \left( 1\cdot 2^E + \beta_{51} \cdot 2^{E-1} + \beta_{50} \cdot 2^{E-2} + \dots + \beta_0\cdot 2^{E-52} \right) \label{eq:binary64} \\ 
\text{with} \quad E = \beta_{62}\cdot 2^{10} + \beta_{61} \cdot 2^9 + \dots + \beta_{52}\cdot 2^0 - 1023. \notag
\end{gather}
A different formula applies if the exponent is \lstinline[language={},mathescape]|0b00$\ldots$00|, to represent numbers close to zero without an implicit leading digit~1. In particular, a 64-bit \lstinline[language={},mathescape]|0b00$\ldots$00| is interpreted as $+0.0$. The exponent \lstinline[language={},mathescape]|0b11$\ldots$11| is used to represent infinite numbers and not-numbers (NaNs). The 4-byte format \lstinline[language=ieee754]|binary32| is defined in an analogous fashion with 8 exponent and 23 significand bits (apart from the leading~1), and exponents ranging from $-126$ to $127$.

On  x86 and amd64 CPUs, real arithmetic is provided by the floating-point instruction sets \emph{x87} and \emph{SSE}. Most of these  instructions expect real numbers to be represented as \lstinline[language=ieee754]{binary32} or \lstinline[language=ieee754]{binary64}, or a sequence of those in a single-instruction-multiple-data (SIMD) vector \cite{amd64}. 
Compilers typically use \lstinline[language=ieee754]|binary32| to implement the C/\Cpp{} type  \lstinline[language=C,keywordstyle=\ttfamily]|float| and the Fortran type \lstinline[language=fortran,keywordstyle=\ttfamily]|real|, and \lstinline[language=ieee754]|binary64| for \lstinline[language=C,keywordstyle=\ttfamily]|double| and \lstinline[language=fortran,keywordstyle=\ttfamily]|double precision|.

Internally, the x87 floating-point registers use an 80-bit double-extended precision format. While the x87 instructions  \lstinline[language={}]|flds|/\lstinline[language={}]|fstps| and \lstinline[language={}]|fldl|/\lstinline[language={}]|fstpl| convert from/to \lstinline[language=ieee754]|binary32| and \lstinline[language=ieee754]|binary64| when they move data between the floating-point registers and memory,  the instructions \lstinline[language={}]|fldt|/\lstinline[language={}]|fstpt| expose the 80-bit format to memory. The GCC and Clang compilers use the 80-bit x87 format to implement the C/\Cpp{} type \lstinline[language=C,keywordstyle=\ttfamily]|long double|.

Our main assumption on the client program is that it  \emph{performs real arithmetic only by the corresponding floating-point instructions} (with the exceptions listed below). To apply forward-mode AD, these instructions have to be recognized and augmented with instructions acting on the operands and their dot values according to basic rules of differential calculus. 

The client program may, moreover, use type-agnostic instructions to move floating-point values, even in several portions like single bytes. In most cases, these instructions are simply applied to the dot values as well. Compare-and-swap (CAS) instructions need special consideration in \cref{sec:cas}.

During our work, we encountered a variety of tricky ways to perform real arithmetic using integer or bitwise logical instructions, or by a ``misuse'' of floating-point instructions. The most important of these \emph{bit-tricks}, presented in \cref{sec:and-or-xor}, can be dealt with by a suitable instrumentation of bitwise logical instructions. Further scenarios listed in \cref{sec:limitations} are not covered by our current implementation. 

\subsection{Compare-And-Swap Instructions}\label{sec:cas}
A CAS instruction like \lstinline|lock cmpxchgq| makes a copy of the present value at a memory location \lstinline[language={}]|addr|, compares it with an ``expected'' value \lstinline|expd|, and only if they match, writes another given value \lstinline|new| to \lstinline[language={}]|addr|. The copy of the previous value at \lstinline[language={}]|addr| can be used to determine whether the write to \lstinline[language={}]|addr| took place.  CAS instructions are useful in setups with multiple threads accessing a shared state, because all of their operations happen in a single ``atomic'' step. Threads can use CAS instructions to make sure in a thread-safe way that when they update the shared state at \lstinline[language={}]|addr|, it has not changed while the update was computed. Otherwise, if a thread $T_1$ used a non-atomic conditional write and another thread $T_2$ wrote to \lstinline[language={}]|addr| after the comparison but before the write of $T_1$, the update by $T_2$ would be discarded.

For the correct AD handling of CAS instructions, it is important to realize that the  shared state of the augmented program consists of both the value at \lstinline[language={}]|addr|, and the dot value in the shadow location. 
Due to their AD augmentation, threads will always update both, but it can happen that an update leaves either the value or the dot value unchanged. To determine whether the shared state has changed, it is therefore mandatory to compare both the present value at \lstinline[language={}]|addr| with the expected value, and the corresponding dot values.
For this reason, we replace a CAS instruction by a construct that first performs the two comparisons, and only if they both yield equality,  writes to memory and to shadow memory.
It is not a problem for  \DerivGrind{} that this construct is non-atomic, because Valgrind runs only one thread at a time and prevents context switches in the middle of an instrumented instruction.

\subsection{Bitwise Logical Instructions}\label{sec:and-or-xor}

\begin{lstfloat}
\centering
\caption{GCC~11.2.0 may use a 128-bit logical ``and'' to set the sign bit of a \lstinline[language=vex]|binary64| to zero, in order to compute the absolute value.}
\label{lst:fabs-and}
\begin{minipage}[t]{0.45\textwidth}
\begin{lstlisting}[language=C,frame=single,showlines=true]
#include <math.h>
double f(double x) {
  return fabs(x); 
}
\end{lstlisting}
\end{minipage}
\quad
\begin{minipage}[t]{0.45\textwidth}
\begin{lstlisting}[language=myasm, frame=single,basicstyle=\ttfamily]
f:
  endbr64
  andpd .LC0(%rip), %xmm0
  ret
; ...
.LC0:
  .long -1         ; 0b11..11
  .long 2147483647 ; 0b01..11
  .long 0          ; 0b00..00
  .long 0          ; 0b00..00
\end{lstlisting}  
\end{minipage}
\end{lstfloat}

A bitwise logical ``and'' operation between a \lstinline[language=ieee754]{binary32}/\lstinline[language=ieee754]{binary64} and \lstinline[language={},mathescape]|0b01$\ldots$11| sets the sign bit to zero, and thus computes the absolute value. \Cref{lst:fabs-and} demonstrates that GCC routinely uses this ``bit-trick''. The assembly code on the right has been produced with GCC~11.2.0 from the C code on the left, using the flags \lstinline[language=bash]|-S| and \lstinline[language=bash]|-O3|. It was stripped from irrelevant labels and directives, and is explained in the following. By the System~V ABI \cite{systemv}, the floating-point argument of the function \lstinline[language=C]|f| is passed in the lower half of the 128-bit register \lstinline|XMM0|. 
The four four-byte constants \lstinline[language={},mathescape]|0b11$\ldots$11|, \lstinline[language={},mathescape]|0b01$\ldots$11|, \lstinline[language={},mathescape]|0b00$\ldots$00| and  \lstinline[language={},mathescape]|0b00$\ldots$00|, inserted by the \lstinline[language={}]|.long| directives, compose the other 128-bit operand to \lstinline[language={}]|andpd|. Note that the most significant byte \lstinline[language={}]|0b01111111| of the second four-byte block  is stored at the highest memory address within that block, as x86 and amd64 stick to the little-endian storage order. Therefore, its zero bit aligns with the sign bit of the \lstinline[language=vex]|binary64| in the lower half of \lstinline|XMM0|.

As a consequence, the AD instrumentation of an ``and'' instruction has to inspect both operands \lstinline[language={}]|x| and \lstinline[language={}]|y|. If \lstinline[language={}]|x|  is equal to \lstinline[language={},mathescape]|0b01$\dots$11|, \lstinline[language={}]|y| might represent a real number whose absolute value is taken by this instruction, and the AD instrumentation must act according to the respective differentiation rule. Analogous considerations apply vice versa.  In case both operands of an ``and'' instruction are \lstinline[language={},mathescape]|0b01$\dots$11|, no differentiable real arithmetic is performed because  \lstinline[language={},mathescape]|0b01$\dots$11|, as a  \lstinline[language=ieee754]{binary32} or \lstinline[language=ieee754]{binary64}, represents not-a-number. 

To handle SIMD operations correctly, the search for \lstinline[language={},mathescape]|0b01$\dots$11| must be performed on all 32-bit and 64-bit blocks of the operands. 

A similar procedure has to be applied for the bitwise logical ``or'' and ``exclusive-or'' instructions when one operand is \lstinline[language={},mathescape]|0b10$\dots$00|, as these operations can compute the negative absolute value or negative value, respectively, of a \lstinline|binary32| or \lstinline|binary64| in the other operand. As an additional complication,  \lstinline[language={},mathescape]|0b10$\dots$00| represents the real number $-0.0$. Therefore, taking the negative of $-0.0$ can lead to an ``exclusive-or'' instruction whose operands  are both  \lstinline[language={},mathescape]|0b10$\dots$00|, making it ambiguous which operand represents the real number, and which dot value the differentiation rule of negation should thus be applied to. 
To resolve this issue, we interpret a logical ``or'' or ``exclusive-or'' as an arithmetic operation only if the dot value of the operand \lstinline[language={},mathescape]|0b10$\dots$00| is \lstinline[language={},mathescape]|0x00$\dots$00|.

\begin{lstfloat}
\centering
\caption{Clang~14.0.0 may use logical operations in a masking pattern to select one of two floating-point numbers.}
\label{lst:clang-masking}
\begin{minipage}[t]{0.45\textwidth}
\begin{lstlisting}[language=C,frame=single,showlines=true]
double f(double a){
  if (a<0){
    return 2+a;
  } else {
    return 2*a;
  }
}
\end{lstlisting}
\end{minipage}
\quad
\begin{minipage}[t]{0.45\textwidth}
\begin{lstlisting}[language=myasm, frame=single,basicstyle=\ttfamily]
.LCPI0_0:
  .quad 0x4000000000000000
f:
  xorpd %xmm1, %xmm1
  movapd  %xmm0, %xmm2
  cmpltsd %xmm1, %xmm2
  movsd .LCPI0_0(%rip), %xmm1
  andpd %xmm2, %xmm1
  andnpd  %xmm0, %xmm2
  orpd  %xmm1, %xmm2
  addsd %xmm2, %xmm0
  retq
\end{lstlisting}  
\end{minipage}
\end{lstfloat}

A second important way  in which floating-point arguments pass through bitwise logical instructions is given by the masking pattern
\begin{center}
\begin{tabular}{c}
\begin{lstlisting}[language=bash,mathescape]
(mask and iftrue) or ((not mask) and iffalse)$.$
\end{lstlisting}
\end{tabular}
\end{center}
The result of this expression is assembled in a bitwise fashion from \lstinline[language={}]|iftrue| and \lstinline[language={}]|iffalse| depending on whether the respective bit in the \lstinline[language={}]|mask| is~1 or~0. We observed this pattern being used by Clang~14.0.0 for the code in \cref{lst:clang-masking}. 
In this example, the mask is created by the  \lstinline[language={}]|cmpltsd| instruction, which sets \lstinline|XMM2| to either  \lstinline[language={},mathescape]|0xff$\ldots$ff| if $a<0$,  or  \lstinline[language={},mathescape]|0x00$\ldots$00| otherwise.
The subsequent ``and'', ``and-not'' and ``or'' instructions then place in \lstinline|XMM2|
either the constant $2.0$ copied from \lstinline[language={}]|.LCPI0_0|, or the value $a$ copied from \lstinline|XMM0|, respectively.
In both cases, finally \lstinline|XMM2| is added to the register \lstinline|XMM0|, which holds the input $a$ in the beginning, and the return value at the end (realizing $2\cdot a$ as $a+a$ in the ``else'' branch).

We support masking patterns as long as entire \lstinline[language=ieee754]{binary32}s or \lstinline[language=ieee754]{binary64}s  are picked in the style of an if-then-else operation, as in the previous example. To this end, \begin{itemize}
\item if one operand of a bitwise logical ``and'' is \lstinline[language={},basicstyle=\ttfamily,mathescape]|0xff$\dots$ff|, or
\item if one operand of a bitwise logical ``or'' is \lstinline[language={},mathescape]|0x00$\dots$00| and has the  dot value \lstinline[language={},mathescape]|0x00$\dots$00|,
\end{itemize}
the dot value of the expression is the dot value of the other operand.

\subsection{Unhandled Hidden Floating-Point Arithmetics}\label{sec:limitations}
Unfortunately, we are not aware of any comprehensive approach to systematically detect all the real arithmetic ``hidden'' in a portion of machine code. This subsection lists scenarios in which our current implementation fails to produce the correct dot values, and indicates patterns that programmers and compilers should avoid to make the machine code ``AD-friendly''.

\paragraph*{Masking of incomplete floating-point representations.} The approach of \cref{sec:and-or-xor} to handle masking patterns breaks down if a \lstinline[language=ieee754]{binary32} or \lstinline[language=ieee754]{binary64} is not entirely selected, e.\,g.\ in order to exchange the exponent bits to implement the C math function \lstinline|frexp|.

\paragraph*{Integer additions to the exponent.} Multiplications with powers of two can be implemented by integer additions to the exponent bits, possibly in conjunction with bit-shifts.   We observed this pattern in the implementation of the exponential function for a \lstinline[language=ieee754]|binary32| SIMD vector in NumPy, where initially  a multiple $k \cdot \ln 2$ is subtracted from the argument to map it into $[0,\ln 2]$, and the result is scaled by $2^k$ to account for this range reduction\footnote{NumPy \cite{numpy} version 1.19.5, \urldefNumPyRange}. It is difficult to decide which of the two summands represents the floating-point number.

\paragraph*{Emulation of floating-point instructions.} As a generalization of the previous items, floating-point libraries like GNU MPFR \cite{mpfr} emulate  arithmetic operations of standardized or proprietary floating-point formats by integer and bitwise logical instructions.    The decimal floating-point arithmetic upcoming with the C\,23 language standard must, as long as there is no hardware support for the IEEE-754 formats \lstinline[language=ieee754]|decimal32|, \lstinline[language=ieee754]|decimal64|, \lstinline[language=ieee754]|decimal128|, also be implemented in software\footnote{GCC 11.2 does it like this: \url{https://github.com/gcc-mirror/gcc/blob/b454c40956947938c9e274d75cef8a43171f3efa/libgcc/config/libbid/bid64_add.c}}.

\paragraph*{Instruction sequences composing a binary identity.} When a client program passes around real numbers as decimal strings or in an otherwise encoded or encrypted form, this obviously erases the dot values. GCC's implementation of an OpenMP atomic addition to a \lstinline[language=C,keywordstyle=\ttfamily]|double|  on the x86 Pentium CPU  initially copies the original value on the stack using the instructions \lstinline[language={}]|fildq|, \lstinline[language={}]|fistpq|, i.\,e.\ reinterprets the value as an integer, converts it to a 80-bit floating-point number, and then converts it back\footnote{GCC 11.2 with \lstinline[language=bash]|-m32 -march=pentium -O3|, \url{https://godbolt.org/z/hc8ehfG88}}.

\paragraph{Rational arithmetic.} As integer operations are ignored by our approach, any calculations with integer fractions are not recognized as performing real arithmetic.

\paragraph{Exploiting floating-point imprecision for rounding.} 
In order to represent a real number $x$ with $2^{52} \leq |x| < 2^{53}$ in the \lstinline[language=ieee754]|binary64| format according to formula~\eqref{eq:binary64}, the exponent $E$ has to be chosen as $52$, so the least-significant bit of the significand controls the binary digit $2^{E-52} = 1$. Therefore within the above range for $x$, the real numbers representable by \lstinline[language=ieee754]|binary64| are precisely the integral numbers. The following procedure exploits this fact to round a \lstinline[language=ieee754]|binary64| $y$ with $|y|< 2^{51}$ to an integral number.
In a first step, $T = 1.5\cdot 2^{52}$ is added to $y$. As $2^{52} < |T+y| < 2^{53}$, storing the sum as a \lstinline[language=ieee754]|binary64| rounds it to an integral number, obeying the floating-point addition's rounding mode. Immediately subtracting $T$ again does not introduce any more floating-point errors, so we end up with the value of $y$ rounded to an integral number. 

The GLIBC math library uses tricks of this kind in several places, e.\,g.\ for range reduction (shifting the argument to $\sin$ into $[-\tfrac{\pi}{4}, \tfrac{\pi}{4}]$ by adding a multiple of $\tfrac{\pi}{2}$)\footnote{GLIBC version 2.35, \urldefGlibcRange} or to compute an index into a lookup table\footnote{GLIBC version 2.35, \urldefGlibcRoundingForTableLookup}.
While the correct dot value of the result of a rounding operation is zero, the instrumented code, according to the above analytical differentiation rules, adds and immediately subtracts $\dot T=0$ from $\dot y$, leaving it unchanged because there are no floating-point errors.

Note that dangerous constructs in the C math library are not problematic for \DerivGrind{} if math function wrappers (\cref{sec:libm-extension}) are used.

\paragraph{Summary and comment.} 
As real arithmetic can be hidden in machine code in various ways, it would be unrealistic to aim for a universal ``AD tool for machine code'' supporting each and every possible client program. Instead, we require that the authors of the client program and the applied compilers stick to the \lstinline[language=ieee754]|binary32|, \lstinline[language=ieee754]|binary64|, and x87 80-bit format to represent real numbers. Furthermore, we assume that they perform real arithmetic only  by the corresponding floating-point instructions, by calls to functions from the C math library, and by the supported bit-tricks listed in \cref{sec:and-or-xor}. From our collection of regression tests described in \cref{sec:application}, only a small number violate this assumption and fail. Therefore, we believe that our assumption is only a mild limitation regarding a productive use of AD for compiled programs.

\section{Implementation of AD Instrumentation in \DerivGrind{}}\label{sec:instrumentation-machine-code}
Instead of directly working with the machine code of the client program, we have implemented the forward-mode AD instrumentation of \cref{sec:ad-forward-theory} using the Valgrind framework \cite{valgrind-paper,valgrind-doc}. Therefore, our tool \DerivGrind{} operates on Valgrind's object-oriented internal representation of machine code, VEX IR, in portions called \emph{superblocks}.

In \cref{sec:vexir}, we revisit the necessary details of VEX. More extensive documentation can be found in the source code of Valgrind\footnote{\urldefValgrindLibvexIR}.  \Cref{sec:shadow} describes how shadow locations for dot values are obtained, and \cref{sec:wrapping-statements,sec:wrapping-expressions} detail how \DerivGrind{} instruments the two main building blocks of VEX to process the dot values. Finally, we remark on system calls in \cref{sec:syscalls} and on limitations of Valgrind in \cref{sec:limitations-valgrind}.

\newcommand{\lrtext}[1]{\langle\text{\rmfamily\itshape #1}\rangle}
\newcommand{\btext}[1]{\text{\rmfamily\itshape #1}}

\subsection{Basic Structure of VEX IR}\label{sec:vexir}
\paragraph*{Synthetic CPU} As VEX IR strives to be independent from hardware and instruction set architectures, its execution model is built around a synthetic CPU. The synthetic CPU has access to memory using the same virtual addresses as the client program. Registers of the synthetic CPU are specified by a byte offset into a separate block of memory called the \emph{guest state}. In addition, VEX IR provides \emph{temporaries} to store intermediate values for the scope of the current superblock. Temporaries are specified by an index and can be assigned only once. 

\paragraph*{Statements and expressions.} Each superblock contains a list of \emph{statements}, which represent actions with side effects, such as those listed in the left column of \cref{tab:vexir-statements}. Depending on the type of action, a statement involves further parameters and \emph{expressions}. A parameter is a value defined during instrumentation, and is denoted by angle brackets $\langle\cdot\rangle$ in \cref{tab:vexir-statements,tab:vexir-expressions}. An expression represents a value to be computed when the containing statement is executed on the synthetic CPU, without side effects. We denote expressions by italic text in \cref{tab:vexir-statements,tab:vexir-expressions,tab:vexir-operations}. The left column of \cref{tab:vexir-expressions} lists types of expressions along with the parameters and sub-expressions that they involve. A very important type of expressions acquire their value on the synthetic CPU by an (integer arithmetic, floating-point arithmetic, bitwise logical, \dots) \emph{operation} applied to one to four sub-expressions. We provide more examples of those in the left column of \cref{tab:vexir-operations}.

\Cref{lst:vex} reproduces an example of a VEX IR superblock's textual representation from the documentation of Valgrind, corresponding to the 4-byte integer addition of the general-purpose register \lstinline|EAX| to \lstinline|EBX| on x86.
Each of the five lines represents one statement. The first statement is meta information, the next three statements write to temporaries and the last statement writes to the guest state. Expressions (indicated by a blue frame) specify which data is written. For the first two writes, the \lstinline|GET| expression represents reads from the registers with byte offsets 0 and 12. The expression on the right hand side of the fourth line applies an operation. Its operands are expressions themselves, which acquire their value by reading from the temporaries with indices 3 and 2. The right hand side of the fifth line  reads from temporary~1.

\begin{lstfloat}
\caption{Example of the textual representation of a VEX IR superblock, from the documentation of Valgrind. Every line represents a statement. We have emphasized parameters with red text and expressions with a blue frame.}
\label{lst:vex}
\vspace{0.5cm}
\framebox{
\parbox{\dimexpr\linewidth-2\fboxsep-2\fboxrule\relax}{
{\ttfamily
-{}-{}-{}-{}-{}- IMark(\vexp{0x24F275}, \vexp{7}, \vexp{0}) -{}-{}-{}-{}-{}-\\
t\vexp{3} = \vexe{GET:\vexp{I32}(\vexp{0})}~~~~~~~~~~~~~\# get \%eax, a 32-bit integer\\
t\vexp{2} = \vexe{GET:\vexp{I32}(\vexp{12})}~~~~~~~~~~~~\# get \%ebx, a 32-bit integer\\
t\vexp{1} = \vexe{\vexp{Add32}(\vexe{t\vexp{3}},\vexe{t\vexp{2}})}~~~~~~~~~~~\# addl\\
PUT(\vexp{0}) = \vexe{t\vexp{1}}~~~~~~~~~~~~~~~~~\# put \%eax
}
}
}
\end{lstfloat}

\paragraph{Dirty calls and CCalls.} VEX achieves its full generality through   \emph{dirty call} statements and  \emph{CCall} expressions. Both store a name, a function pointer and a tuple of expressions (among other things), and when execution on the synthetic CPU reaches the dirty call or CCall, the stored function is called with arguments taken from the evaluated expressions. The Valgrind core represents certain x86 and amd64  instructions  by dirty calls when there is no architecture-independent equivalent in VEX. 

For example, the x87 instruction \lstinline[language={}]|fldt| loads 10~bytes of memory into an x87 floating-point register, which internally uses 80~bits to represent  a floating-point number. The other way round, \lstinline[language={}]|fstpt| stores an 80-bit floating-point representation in memory. Note that VEX IR does not support 80-bit arithmetic.  Instead,  Valgrind represents the x87 floating-point registers by \lstinline[language=ieee754]|binary64|s in the guest state of the synthetic CPU, and replaces 80-bit by 64-bit arithmetic.  Therefore, dirty calls replacing  the \lstinline[language={}]|fldt| and \lstinline[language={}]|fstpt| instructions have to convert between the \lstinline[language=ieee754]{binary64} format in the guest state and the 80-bit format in memory.

Additionally, \DerivGrind{} emits dirty calls to access the shadow memory from within VEX, and CCalls for the rather complicated AD augmentation of bitwise logical instructions (\cref{sec:and-or-xor}).

\subsection{Shadow Temporaries, Registers and Memory}\label{sec:shadow}
As described in \cref{sec:shadowing}, \DerivGrind{} provides a shadow location to every storage location.
\paragraph*{Temporaries} Within each superblock,   \DerivGrind{} shadows every temporary $i$ with the temporary $(i+m_\text{tmp})$, using a shift $m_\text{tmp}$ larger than the maximal index of a temporary before instrumentation. When \DerivGrind{} needs additional temporaries to compute dot values, it allocates them with indices greater than or equal to $2 \cdot m_\text{tmp}$. 

\paragraph*{Registers} Before instrumentation, the VEX code uses a known, architecture-dependent number $m_{\text{gs}}$ of bytes of the guest state. Valgrind's synthetic CPU provides $3 \cdot m_\text{gs}$ bytes of guest state to the instrumented code. For each register with byte offset $j$, \DerivGrind{} can therefore use the ``shadow register'' with byte offset $j + m_\text{gs}$ to store the dot value. 

\paragraph*{Memory} The address shifting approach used to shadow temporaries and registers is not feasible for memory, as the client program can use addresses throughout the whole virtual address space, and the tool has limited control about which parts of the virtual address space it might itself make allocations in. Like Memcheck, \DerivGrind{} therefore uses a \emph{shadow memory tool}, developed by us\footnote{\urldefShadowmemGithub} according to \citet{shadowmem-nethercote}. It can be thought of as a big hashmap assigning the content of shadow memory to memory addresses; the actual implementation uses a prefix tree (trie) data structure similar to multilevel page tables for virtual memory. As the default value of an uninitialized byte in the shadow memory is \lstinline[language={}]|0x00|,   \lstinline[language=ieee754]|binary32|s and \lstinline[language=ieee754]|binary64|s in the data and bss segment of the client program (such as C/\Cpp{} floating-point variables with static storage duration)  have the correct initial dot value~$+0.0$.

\subsection{Wrapping Statements}\label{sec:wrapping-statements}

\begin{table}
\caption{VEX IR statements, and their augmentation with forward-mode AD logic.}
\label{tab:vexir-statements}
\begin{tabular}{p{0.48\textwidth}p{0.48\textwidth}}
\toprule
VEX statement & Differentiated VEX statement \\
\midrule
Write data to a temporary with index $i$. & Write differentiated data to the shadow temporary. \\
\lstinline[language=vex,mathescape]|t$\langle i\rangle$ = $p$| & 
\lstinline[language=vex,mathescape]|t$\langle i+m_\text{\rmfamily tmp}\rangle$ = $\dot p$| 
\\
\midrule
Write data to the register with byte offset $j$ in the guest state. & Write differentiated data to the shadow register. \\
\lstinline[language=vex,mathescape]|PUT($\langle j \rangle$) = $p$| & 
\lstinline[language=vex,mathescape]|PUT($\langle j + m_\text{\rmfamily gs} \rangle $) = $\dot p$| 
\\
\midrule
Write data to memory. \hfill\hfill\hfill\linebreak
\lstinline[language=vex,mathescape]|STle($\btext{address}$) = $p$| & 
Write differentiated data to shadow memory. Implemented as a dirty call to access the shadow memory from VEX. 
\\
\midrule
Write data to memory, if a condition is satisfied. \hfill\hfill\hfill\linebreak
\lstinline[language=vex,mathescape]|if ($\btext{guard}$) STle($\btext{address}$) = $p$| & 
 Write differentiated data to shadow memory, subject to the same condition. Implemented as a dirty call. 
\\
\midrule
Compare-and-swap, loading $\btext{addr}$ into temporary \lstinline[language=vex,mathescape]|t$\lrtext{old}$|, and replacing data at \emph{addr} by $\btext{new}$ if it matches $\btext{expd}$. & CAS statements are replaced as discussed in \cref{sec:cas}.\\
\lstinline[language=vex,mathescape]|t$\lrtext{old}$ = CASle($\btext{addr}$ :: $\btext{expd}$ -> $\btext{new}$)| & 
\\
\midrule
Dirty call, invoking a Valgrind function with side effects. & The augmentation depends on the dirty call, see details in \cref{sec:wrapping-statements}.  \\
\midrule
Meta information. & Not relevant for AD. \\
\lstinline[language=vex,mathescape]|------ IMark($\dots$) ------| & 
\\
\midrule
Conditional jump. & Not relevant for AD. \\
\lstinline[language=vex,mathescape]|if ($\btext{guard}$) goto {$\lrtext{jump kind}$} $\lrtext{target}$| &
\\
\bottomrule
\end{tabular}
\end{table}

For all statements with relevance to AD and except for CAS statements, \DerivGrind{} inserts the forward-mode AD logic \eqref{eq:forward-update} in the form of a ``differentiated statement'' in front of the original statement. Examples of such differentiated statements are shown in \cref{tab:vexir-statements}. Most of them involve ``differentiated expressions'' that compute the dot value of a value computed by an expression in the original statement. Differentiated expressions have been marked with a dot, and can be formed as described in \cref{sec:wrapping-expressions}. 

\DerivGrind{} replaces a CAS statement by an entirely new sequence of statements, in order to perform the additional check whether the dot value has changed, as discussed in \cref{sec:cas}. 

Dirty call statements can be identified by their name. \DerivGrind{} matches dirty calls emitted for \lstinline[language={}]|fldt| and \lstinline[language={}]|fstpt| (see \cref{sec:vexir}) by dirty calls  performing the same operation on the shadow guest state and shadow memory, to comply with our convention to store the dot value of every floating-point variable in the same format as its value (\cref{sec:shadowing}).

As far as we observed it, all the other dirty calls emitted by the Valgrind core from x86 and amd64 machine code can hardly be part of a floating-point calculation. For example, they perform actions such as obtaining information about the CPU hardware (the VEX IR equivalent of the x86 instruction \lstinline[language={}]|cpuid|), reading a time-stamp counter (\lstinline[language={}]|rdtsc|), SSE~4.2 string operations (\lstinline[language={},mathescape]|pcmp$\langle X\rangle$str$\langle Y\rangle$|), or (re)storing some SSE status bits. These dirty calls do not require any additional AD logic, except that shadows of output temporaries should be assigned some value so they are initialized in case they are later read from.

\subsection{Wrapping Expressions}\label{sec:wrapping-expressions}

\begin{table}
\caption{VEX IR expressions, and their forward-mode algorithmic derivatives.}
\label{tab:vexir-expressions}
\begin{tabular}{p{0.48\textwidth}p{0.48\textwidth}}
\toprule
Expression $p$ & Differentiated expression $\dot p$ \\
\midrule
Read from a temporary with index $i$. & Read from the shadow temporary. \\
\lstinline[language=vex,mathescape]|t$\langle i \rangle$| &
\lstinline[language=vex,mathescape]|t$\langle i + m_\text{\rmfamily tmp} \rangle$| 
\\
\midrule
Read data of specified type from the register with byte offset $j$ in the  guest state. & Read data of the same type from the shadow register. \\
\lstinline[language=vex,mathescape]|GET:$\lrtext{type}$($\langle j \rangle $)| &
\lstinline[language=vex,mathescape]|GET:$\lrtext{type}$($\langle j + m_\text{\rmfamily gs} \rangle$)| 
\\
\midrule
Read from memory. \hfill\hfill\hfill\linebreak
\lstinline[language=vex,mathescape]|LDle:$\lrtext{type}$($\btext{address}$)| &
 Read from shadow memory, expecting data of the same type. Implemented as a dirty call.
\\
\midrule
Operation with one to four arguments, see \cref{tab:vexir-operations} for examples. & See \cref{tab:vexir-operations}. \\
\lstinline[language=vex,mathescape]|$\lrtext{op}$($q_1$, $q_2$, $\dots$ )| &
\\
\midrule
Constant value. & Constant value zero of the same type. \\
\lstinline[language=vex,mathescape]|$\lrtext{literal}$:$\lrtext{type}$| or 
\lstinline[language=vex,mathescape]|$\lrtext{type}${$\lrtext{literal}$}| &
\lstinline[language=vex,mathescape]|0x0:$\lrtext{type}$| or
\lstinline[language=vex,mathescape]|$\lrtext{type}${0x0}| 
\\
\midrule
If-then-else construct, selecting either \lstinline[mathescape]|$q_\text{true}$| or  \lstinline[mathescape]|$q_\text{false}$| depending on $\btext{condition}$. & If-then-else construct with the same condition on differentiated operands. \\
\lstinline[language=vex,mathescape]|ITE($\btext{condition}$, $q_\text{true}$, $q_\text{false}$)| &
\lstinline[language=vex,mathescape]|ITE($\btext{condition}$, $\dot q_\text{true}$, $\dot q_\text{false}$)| 
\\
\midrule
CCall to Valgrind function without side effects. & Until now, we only encountered cases without relevance to AD. 
\\
\bottomrule
\end{tabular}
\end{table}

\begin{table}
\caption{VEX IR expressions performing an operation, and their forward-mode algorithmic derivatives. The placeholder \emph{rm} represents an expression for the rounding mode.}
\label{tab:vexir-operations}
\begin{tabular}{p{0.48\textwidth}p{0.48\textwidth}}
\toprule
Expression $p$ & Differentiated expression $\dot p$ \\
\midrule
Scalar floating-point arithmetic, e.\,g.\ addition of \lstinline[language=ieee754]|binary64|s, & Application of the differentiation rule. \\
\lstinline[language=vex,mathescape]|AddF64($rm$,$q$,$s$)| &
\lstinline[language=vex,mathescape]|AddF64($rm$,$\dot q$,$\dot s$)| \\
or multiplication of \lstinline[language=ieee754]|binary32|s, &  \\
\lstinline[language=vex,mathescape]|MulF32($rm$,$q$,$s$)| &
\lstinline[language=vex,mathescape]|AddF32($rm$,MulF32($rm$,$\dot q$,$s$),|
\phantom{\lstinline[language=vex,mathescape]|AddF32($rm$,|}\lstinline[language=vex,mathescape]|MulF32($rm$,$q$,$\dot s$))|
\\
\midrule
SIMD floating-point arithmetic, e.\,g.\ multiplication of eight \lstinline[language=ieee754]|binary32|s. & Component-wise application of the differentiation rule. \\
\lstinline[language=vex,mathescape]|Mul32Fx8($rm$,$q$,$s$)| &
\lstinline[language=vex,mathescape]|Add32Fx8($rm$,Mul32Fx8($rm$,$\dot q$,$s$),| 
\phantom{\lstinline[language=vex,mathescape]|Add32Fx8($rm$,|}\lstinline[language=vex,mathescape]|Mul32Fx8($rm$,$q$,$\dot s$))|
\\
\midrule
Lowest-lane-only SIMD floating-point arithmetic, e.\,g.\ mapping the operands $(q_0,q_1,q_2,q_3)$ and $(s_0,s_1,s_2,s_3$) to $(q_0 \cdot s_0, q_1,q_2,q_3)$. & Formal application of the differentiation rule with lowest-lane-only operations, taking care to be correct outside the lowest lane also. E.\,g.\ for $(\dot q_0 s_0 + q_0 \dot s_0, \dot q_1, \dot q_2, \dot q_3)$,  \\
\lstinline[language=vex,mathescape]|Mul32F0x4($q$,$s$)| &
\lstinline[language=vex,mathescape]|Add32F0x4(Mul32F0x4($\dot q$,$s$),|
\phantom{\lstinline[language=vex,mathescape]|Add32F0x4(|}\lstinline[language=vex,mathescape]|Mul32F0x4($q$,$\dot s$))|
\\
\midrule
Floating-point conversions, e.\,g. & Analogous application to the dot values. \\
\lstinline[language=vex,mathescape]|F64toF32($rm$,$q$)| &
\lstinline[language=vex,mathescape]|F64toF32($rm$,$\dot q$)|
\\
\midrule
Binary reinterpretation of floating-point representations as integers and vice versa, e.\,g. & Analogous application to the dot values. \\
\lstinline[language=vex,mathescape]|ReinterpI64asF64($q$)| &
\lstinline[language=vex,mathescape]|ReinterpI64asF64($\dot q$)| 
\\
\midrule
SIMD (un)packing, e.\,g. & Analogous application to the dot values. \\
\lstinline[language=vex,mathescape]|64x4toV256($q_3$,$q_2$,$q_1$,$q_0$)| &
\lstinline[language=vex,mathescape]|64x4toV256($\dot q_3$,$\dot q_2$,$\dot q_1$,$\dot q_0$)|
\\
\midrule
Bitwise logical operations, e.\,g. & Handling according to \cref{sec:and-or-xor}. \\
\lstinline[language=vex,mathescape]|And64($q$,$s$)| &
(Represented by a CCall.)
\\
\midrule
Integer arithmetic, e.\,g. & Not relevant for AD. \\
\lstinline[language=vex,mathescape]|Add64($q$,$s$)| &
\lstinline[language=vex]|0x0:I64|
\\
\midrule
Comparisons, e.\,g. & Not relevant for AD. \\
\lstinline[language=vex,mathescape]|CmpF64($q$,$s$)| &
\lstinline[language=vex]|0x0:I32|
\\
\bottomrule

\end{tabular}
\end{table}

Many of the differentiated statements (\cref{sec:wrapping-statements} and \cref{tab:vexir-statements}) involve a differentiated expression $\dot p$ computing the dot value of the result of an expression $p$ in the original statement. \DerivGrind{} forms differentiated expressions according to \cref{tab:vexir-expressions}. \Cref{tab:vexir-operations} gives more details for the important subclass of expressions that perform an  operation. From the large number of operations available in VEX, we only handle those that we consider necessary and that we suspect to be covered by our regression tests (\cref{sec:unittests}). For instance, the VEX operation \lstinline[language=vex]|SinF64|, corresponding to the x87 instruction \lstinline[language={}]|fsin|, is currently not handled because apparantly, modern compilers and libraries do not use this instruction. For unhandled operations, \DerivGrind{} assumes a zero derivative, and optionally outputs the containing statement for debugging purposes.

\subsection{System Calls}\label{sec:syscalls}
Under POSIX-compliant operating systems, userspace programs accomplish (most of) their interaction with the outside by invoking functionality of the kernel.  Such \emph{system calls} are usually raised through software interrupts or specific instructions. Valgrind does not instrument kernel code, but enables tools to wrap system calls. 

For now, \DerivGrind{} does not use this feature. Therefore, writing data to standard, file or network streams does not export the dot values. Data read from streams into memory is endowed with the dot values previously residing in the corresponding parts of the shadow memory. In particular, \DerivGrind{} currently cannot handle MPI multiprocessing.

\subsection{Limitations of Valgrind}\label{sec:limitations-valgrind}
From the limitations listed in the Valgrind documentation \cite{valgrind-doc}, the following have a particular relevance for AD.
\begin{itemize}
\item Valgrind does not support 3DNow!\ instructions (which are anyway rarely supported by CPUs) and AVX-512 instructions (see \cite{avx512-in-valgrind} for current development efforts). When Valgrind encounters an unknown instruction, it sends SIGILL to the client program, triggering its termination if it did not define a signal handler. 
\item Valgrind replaces 80-bit by 64-bit floating-point arithmetic (as mentioned above), and performs these with partial observance of rounding modes, no support for numeric exceptions, and ignoring some SSE2 control bits. If the client program is very sensitive to floating-point errors, it might therefore behave differently when executed under Valgrind. We do not expect this to become a problem, since algorithmic derivatives of very sensitive programs are usually not meaningful.
\end{itemize}

\section{Accessing Dot Values in \DerivGrind{}}\label{sec:interaction}
\DerivGrind{}'s instrumentation of the machine code propagates dot values alongside the execution of the compiled client program. This part of the tool, as described in \cref{sec:ad-forward-theory,sec:instrumentation-machine-code}, does not rely on the source of the client program.

However, the user needs some knowledge on the internal structure of the client program in order to identify input and output variables for AD. In order to \emph{seed} dot values of inputs before the propagation, and retrieve dot values of outputs afterwards, these  variables must be matched to memory addresses. It is therefore natural that \DerivGrind{}'s interfaces for setting and getting dot variables rely on the source code of the client program to some extent. We describe two interfaces in this section.

\subsection{Monitor Commands Interface}\label{sec:interaction-monitor}
Valgrind's \emph{monitor commands} mechanism enables the user to interact with Valgrind during the execution of the client program. When Valgrind is started with the command-line argument \lstinline[language=bash]|--vgdb-error=0|, it activates its built-in gdbserver and waits for a connection, instead of executing the instrumented client program right away.  The user has to connect to the gdbserver from a GDB session with GDB's \lstinline[language=bash]|target remote| command.  In addition to regular debugger commands like setting breakpoints, stepping, and inspecting memory, the user can then send \emph{monitor commands} over this connection. \DerivGrind{} provides monitor commands to access the shadow memory, and hence the dot value of any variable. Therefore, this mechanism allows for an interactive exploration of automatic derivatives of any variable at any point of time, and with respect to any variable at any (earlier) point of time, during the execution of a client program---as long as the debugger can stop the client  at these points of time, and the user can obtain memory addresses of the variables.

This condition implies that the source files which either define variables of interest, or contain lines where a breakpoint should be set, can be read by the user, and recompiled with debugging symbols (e.\,g.\ \lstinline[language=bash]|-g| flag of GCC and Clang) and most optimizations  turned off (e.\,g.~\lstinline[language=bash]|-O0|). 

\subsection{Valgrind Client Request Interface}\label{sec:interaction-clientrequest}
Valgrind's \emph{client request} mechanism enables the client program  to interact with the Valgrind core and tool. 
In contrast to the monitor commands interface (\cref{sec:interaction-monitor}) where the user selects and accesses AD input and output variables interactively in a debugger, the client request interface enables the user to define them by code inserted into the client program.

To perform a request, the client program has to assemble a data structure specifying the request, load its address into a specific register, and then execute a specific sequence of machine code instructions. On a normal CPU, this instruction sequence amounts to a no-operation.  When the client is running under Valgrind however, Valgrind recognizes the pattern and passes the data structure to client request handlers in the Valgrind core and tool. It is easy to make a client request from an editable C/\Cpp{} source, because Valgrind provides header files with preprocessor macros that set up the data structure and add the specific instruction sequence using the \lstinline[language=C]|__asm__| syntax.

\DerivGrind{} defines and implements client requests to copy data from memory to shadow memory and vice versa. \Cref{lst:clientrequest} demonstrates the usage of the corresponding macros, which are called with a memory address, a shadow memory address, and the number of bytes to be copied. Note that when function arguments  are passed by value, the AD instrumentation makes sure that their dot values are copied as well. The dot value of \lstinline[language=C]|dotvalue| is irrelevant for the setter, and unspecified for the getter.

\begin{lstfloat}
\caption{Usage of \DerivGrind{}'s client request macros to access the shadow memory from within the client program.}
\label{lst:clientrequest}
\vspace{0.5cm}
\begin{minipage}[t]{\textwidth}
\begin{lstlisting}[language=C,frame=single,showlines=true,mathescape]
#include <valgrind/derivgrind.h>
double set_dotvalue(double value, double dotvalue){
  // Copy 8 bytes from the memory address &dotvalue 
  // to the shadow memory address &value.
  DG_SET_DOTVALUE(&value, &dotvalue, sizeof(double));
  return value;
}
double get_dotvalue(double value){
  double dotvalue = 0.; // Return 0. if run outside Valgrind.
  // Copy 8 bytes from the shadow memory address &value
  // to the memory address &dotvalue.
  DG_GET_DOTVALUE(&value, &dotvalue, sizeof(double));
  return dotvalue;
}
\end{lstlisting}
\end{minipage}
\end{lstfloat}

A user of the client request interface has to insert calls to the C functions \lstinline[language=C]|set_dotvalue| (typically proceeded by an initialization of the dot value) and \lstinline[language=C]|get_dotvalue| (typically followed by an output statement) near the lines of the client's source code where the input and output variables are defined. It is therefore necessary that the respective parts of the source code can be edited and recompiled, and that the respective programming languages and compilers provide a ``C interface''. In the special case of differentiating a Python interpreter in \cref{sec:result-python}, we are able to make client requests without any modifications of the interpreter's source code.

\section{Wrapping the C Math Library}\label{sec:libm-extension}

\begin{figure}
\centering
\begin{tikzpicture}
\begin{axis}[xlabel={$x$},ylabel style={align=center},width=0.5\textwidth,ymax=1.5]
\addplot[black,very thick] table[x index=0,y index=2] {images/sin-without-vgpreload.dat};
\addlegendentry{$\frac{\mathrm d}{\mathrm dx} \text{{\ttfamily sin}}(x)$}
\addplot[black,very thin,dashed] table[x index=0,y index=3] {images/sin-without-vgpreload.dat};
\addlegendentry{$\text{{\ttfamily cos}}(x)$}
\end{axis}
\end{tikzpicture}
\hfill
\begin{tikzpicture}
\begin{axis}[xlabel={$x$},ylabel style={align=center}, width=0.5\textwidth]
\addplot[black,very thick] table[x index=0,y index=2] {images/log-without-vgpreload.dat};
\addlegendentry{$\frac{\mathrm d}{\mathrm dx} \text{{\ttfamily log}}(x)$}
\addplot[black,very thin,dashed] table[x index=0,y index=3] {images/log-without-vgpreload.dat};
\addlegendentry{{\ttfamily 1/x}}
\end{axis}
\end{tikzpicture}

\caption{The black-box algorithmic derivative of GLIBC's implementation of \lstinline[language=C]|sin| and \lstinline[language=C]|log|  agrees with the analytic derivatives only within some intervals.}
\label{fig:deriv-without-vgpreload}
\Description{Two plots comparing the black-box derivative of sin and log with the analytic derivative.}
\end{figure}

The C standard library provides basic maths functions such as power and square root, the trigonometric and hyperbolic functions and their inverses,  exponentiation  and logarithm. While some of them could be realized by hardware instructions like  \lstinline[language={},keywordstyle=\ttfamily]|fsin| and \lstinline[language={},keywordstyle=\ttfamily]|fcos|, implementations of the standard library are free to perform an approximation algorithm entirely in software. 

\Cref{fig:deriv-without-vgpreload} shows the derivatives of the GLIBC math library's implementation of \lstinline[language=C]|sin| and \lstinline[language=C]|log|,  using the components of \DerivGrind{} presented so far. The algorithmic derivatives match the analytic derivatives $\cos(x)$ and $1/x$ only inside the intervals $[-0.126, 0.126]$ and $[0.9375,1.0646972656\dots]$, respectively,  and are zero outside. We further analyzed the case of \lstinline[language=C]|sin| with the following findings:
\begin{itemize}
\item For $|x|<2^{-26}$ and $2^{-26}\leq |x|<0.126$, \lstinline[language=C]|sin(x)| is computed using the Taylor polynomials of degree~1 or~11, respectively. \DerivGrind{} thus computes the derivatives of these polynomials, which equal the Taylor polynomials of degree~0 or~10 to the cosine function, and are therefore good approximations for the analytical derivative in the respective intervals.
\item For $0.126\leq |x| < 0.855\dots$, the algorithm in GLIBC is based on a trigonometric formula \begin{equation*}
\sin(x_\text{tab} + x_\text{rem}) = \sin(x_\text{tab}) \cos(x_\text{rem}) + \cos(x_\text{tab}) \sin(x_\text{rem})
\end{equation*}
after writing $x$ as $x_\text{tab}+x_\text{rem}$ with a multiple $x_\text{tab}$ of $2^{-7}$ and a small remainder $x_\text{rem}$. The purpose of this decomposition is to read the sine and cosine  of $x_\text{tab}$ from a lookup table, and to use a Taylor series  for $x_\text{rem}$. While the correct decomposition of $\dot x$ would be $\dot x_\text{tab} = 0$ and $\dot x_\text{rem} = \dot x$,  \DerivGrind{} erroneously computes $\dot x_\text{tab}=\dot x$ and $\dot x_\text{rem} = 0$ because GLIBC performs the decomposition by adding and subtracting a big constant, relying on floating-point errors as described in \cref{sec:limitations}.
\item For $0.855\ldots \leq |x| < 2.426\dots$, the implementation of \lstinline[language=c]|cos| is invoked with a modified value, basically using the same lookup-table based approach. 
\item For $2.426\ldots < |x| < 1.054\ldots\cdot 10^8$, the previously mentioned methods are used for a shifted argument $y = x - k\cdot \tfrac{\pi}{2} \in [-\tfrac{\pi}{4}, \tfrac{\pi}{4}]$. As GLIBC again computes the integral factor $k$ by a tricky exploitation of floating-point errors as described in \cref{sec:limitations}, \DerivGrind{} erroneously finds $\dot k = \dot x \cdot \tfrac{2}{\pi}$ and $\dot y = 0$. 
\end{itemize}

Fortunately, the Valgrind function wrapping feature allows Valgrind tools to  specify functions by their names (and, if desired, the ``soname'' field of the containing shared object) and reroute the respective calls to wrapper functions supplied by the tool. \DerivGrind{} wraps all C\,95 math functions using this mechanism. The wrappers obtain their arguments' dot values by the client request mechanism (\cref{sec:interaction-clientrequest}), compute the return value and its analytical derivative using the original math functions, and overwrite the return value's dot value accordingly using the client request mechanism.

Our approach is not universal. If a client program implements numerical approximations of mathematical functions on its own and uses different function names or inlining, AD tools based on machine code can hardly recognize these, and thus fall back to black-box differentiation. For instance, \DerivGrind{} computes wrong derivatives in several testcases involving NumPy's 32-bit floating-point type, because NumPy reimplements some  math functions for \lstinline[language=ieee754]|binary32| on amd64. And the other way round, if a client program reuses \lstinline[language=bash]|math.h| function names in a shared object \lstinline|libm.so*| with a different semantic or signature, it might cause unexpected behaviour.

\section{Evaluation}\label{sec:application}
Our collection of regression tests, described in \cref{sec:unittests}, checks \DerivGrind{}'s results for a large number of small pieces of code.  We apply \DerivGrind{} to larger programs in \cref{sec:result-performance,sec:result-python} for further validation, and  to measure the average increase in time and memory complexity. In \cref{sec:result-performance}, the differentiated program is a numerical solver for Burgers' partial differential equation (PDE). In \cref{sec:result-python}, we differentiate a Python interpreter, as an example for a large pre-compiled program whose source code is not available to \DerivGrind{}.

\subsection{Regression Tests}\label{sec:unittests}
Our test suite verifies the values and derivatives computed by \DerivGrind{} for many combinations of \begin{itemize}
\item a language and compiler: C and \Cpp{} programs are compiled by GCC and Clang, Fortran programs are compiled by GCC; Python scripts are interpreted by CPython (see more details in \cref{sec:result-python}),
\item a simple ``algorithm'' to be differentiated:  elementary operations, calls to \lstinline[language=bash]|math.h| functions, control structures, loops suitable for auto-vectorization, and OpenMP constructs;
\item a floating-point type: \lstinline[language=ieee754]|binary32|, \lstinline[language=ieee754]|binary64|, and the 80-bit x87 double-extended precision type; and
\item an architecure: x86 or amd64.
\end{itemize}
\DerivGrind{} passes almost all of these tests, demonstrating its versatility in general.
The small number of failing tests were either related to applying NumPy math functions on  \lstinline[language=ieee754]|binary32| arguments in CPython on amd64, or to using OpenMP constructs with GCC.  The underlying issues are discussed in \cref{sec:limitations,sec:libm-extension} in more detail.

\subsection{Performance Study}\label{sec:result-performance}
We apply \DerivGrind{} to a PDE solver for the two-dimensional Burgers' equations on a unit square, in order to differentiate a norm of the solution after the last timestep with respect to a simultaneous shift of all components of the initial state. Arithmetically, the \Cpp{} code merely involves addition, subtraction, multiplication, division, and the square root function. A related benchmark has been used in previous studies of AD performance \cite{sagebaum_expression_2018,SaAlGauTOMS2019,sagebaum2021index,BluehdornSG2021}. For all the configurations considered in the following, the derivatives computed by \DerivGrind{} match those of CoDiPack's forward mode. 

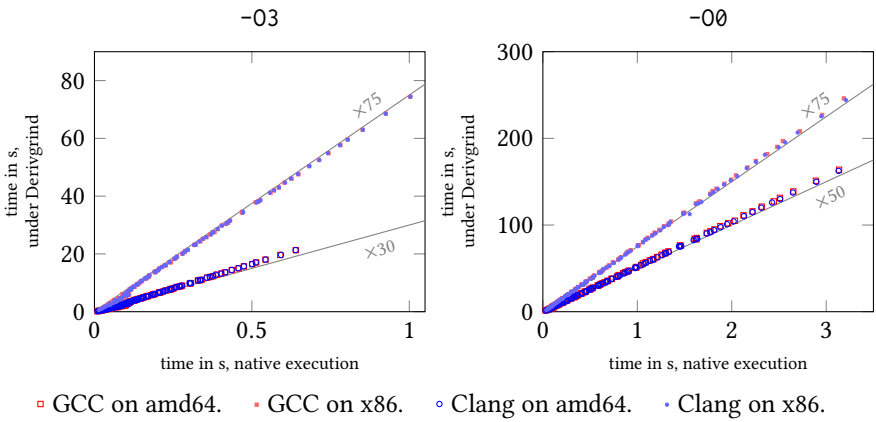
\begin{figure}
\centering
\begin{tikzpicture}
\begin{axis}[title={{\ttfamily -O3}},xlabel={time in s, native execution}, ylabel={time in s, \\ under \DerivGrind{}},xlabel style={align=center,font=\scriptsize},ylabel style={align=center,font=\scriptsize}, width=0.43\textwidth, height=5cm, xmin=0,xmax=1.05,ymin=0,ymax=90]
\addplot[red,mark=square,only marks,mark size=1pt] table[x index=1, y index=3] {images/ftimeram2/PERF_OUTPUT_amd64_g++_o3_awk};
\addplot[red!60,mark=square*,only marks,mark size=0.6pt] table[x index=1, y index=3] {images/ftimeram3/PERF_OUTPUT_x86_g++_o3_awk};
\addplot[blue,mark=o,only marks,mark size=1.0pt] table[x index=1, y index=3] {images/ftimeram2/PERF_OUTPUT_amd64_clang++_o3_awk};
\addplot[blue!60,mark=*,only marks,mark size=0.6pt] table[x index=1, y index=3] {images/ftimeram3/PERF_OUTPUT_x86_clang++_o3_awk};
\draw [gray,thin] (axis cs: 0,0) -- (axis cs: 1.05,31.5) node[pos=0.85,sloped,below] {\scriptsize $\times 30$};
\draw [gray,thin] (axis cs: 0,0) -- (axis cs: 1.05,78.75) node[pos=0.85,sloped,above] {\scriptsize $\times 75$};
\end{axis}
\end{tikzpicture}
\begin{tikzpicture}
\begin{axis}[title={{\ttfamily -O0}},xlabel={time in s, native execution}, ylabel={time in s, \\ under \DerivGrind{}},xlabel style={align=center,font=\scriptsize},ylabel style={align=center,font=\scriptsize}, width=0.43\textwidth, height=5cm, xmin=0,xmax=3.5,ymin=0,ymax=300]
\addplot[red,mark=square,only marks,mark size=1pt] table[x index=1, y index=3] {images/ftimeram2/PERF_OUTPUT_amd64_g++_o0_awk};
\addplot[red!60,mark=square*,only marks,mark size=0.6pt] table[x index=1, y index=3] {images/ftimeram3/PERF_OUTPUT_x86_g++_o0_awk};
\addplot[blue,mark=o,only marks,mark size=1pt] table[x index=1, y index=3] {images/ftimeram2/PERF_OUTPUT_amd64_clang++_o0_awk};
\addplot[blue!60,mark=*,only marks,mark size=0.6pt] table[x index=1, y index=3] {images/ftimeram3/PERF_OUTPUT_x86_clang++_o0_awk};
\draw [gray,thin] (axis cs: 0,0) -- (axis cs: 3.5,175.0) node[pos=0.85,sloped,below] {\scriptsize $\times 50$};
\draw [gray,thin] (axis cs: 0,0) -- (axis cs: 3.5,262.5) node[pos=0.85,sloped,above] {\scriptsize $\times 75$};
\end{axis}
\end{tikzpicture}
\hfill \\
    \raisebox{0.20em}{\addlegendimageintext{red,mark=square,only marks,mark size=1pt}} GCC on amd64. \quad
    \raisebox{0.20em}{\addlegendimageintext{red!60,mark=square*,only marks,mark size=0.6pt}} GCC on x86. \quad
     \raisebox{0.20em}{\addlegendimageintext{blue,mark=o,only marks,mark size=1pt}} Clang on amd64. \quad
     \raisebox{0.20em}{\addlegendimageintext{blue!60,mark=*,only marks,mark size=0.6pt}} Clang on x86. \quad
\caption{\DerivGrind{}'s effect on the run-time of the Burgers benchmark.}
\label{fig:scalingfactor-time}
\Description{Two scatter plots titled -O3, -O0, with x-axis label ``time in s, native execution'' and y-axis label ``time in s, under Derivgrind''.}
\end{figure}

\Cref{fig:scalingfactor-time} displays the effect of \DerivGrind{} on the run-time. We considered~$2^3=8$ setups, using the GCC~10.2.1 (\lstinline[language=bash]|g++|) and Clang~11.0.1 (\lstinline[language=bash]|clang++|) compilers, for amd64 (no flag) and x86 (\lstinline[language=bash]|-m32|), with full (\lstinline[language=bash]|-O3|) and without (\lstinline[language=bash]|-O0|) optimization. 
Our time measurements refer to the difference in the system time retrieved by the client program right before and after solving the PDE.  This eliminates the constant startup and finalization time of \DerivGrind{} and the shadow memory tool, which may take up to around \SI{2}{\second} depending on the configuration. Averages were taken over~100 (\lstinline[language=bash]|-O3|) or~10 (\lstinline[language=bash]|-O0|) measurements. 
The client program was compiled and executed on an exclusive 64-bit Intel Xeon Gold~6126 processor at \SI{2.6}{\giga\hertz} in the Elwetritsch cluster at the University of Kaiserslautern-Landau. 

Each dot in \cref{fig:scalingfactor-time} represents a problem instance with an $n_x \times n_x$ grid and $n_t$ time steps, for $n_x=100, 120, \dots, 500$ and $n_t=100,200,\dots,500$.  
The plots show that \DerivGrind{} slows down the PDE solver by a factor that is essentially independent from $n_x$ and $n_t$, and varies between~30 and~75. The best factor of about~30 is reached for the practically most relevant case of an optimized build on amd64. As \DerivGrind{}'s instrumentations of the various VEX constructs differ in complexity, and probably offer a different amount of opportunities for optimizations by the Valgrind core, it is natural that the slow-down factor depends on the ``mixture'' of instructions produced by the compiler. 
For comparison, when running the benchmark with the forward mode of the AD tool CoDiPack, the largest slow-down factor measured by us on the setups with \lstinline[language=bash]|-O3| is approximately~$3.3$. 

\begin{figure}
\centering
\begin{tikzpicture}
\begin{axis}[title={},xlabel={maximum resident set size in KB, native execution}, ylabel={max.\ RSS in KB,\\ under \DerivGrind{}},xlabel style={align=center,font=\scriptsize},ylabel style={align=center,font=\scriptsize}, width=0.9\textwidth, height=5cm, xmin=0,xmax=1200000,ymin=0,ymax=7000000]
\addplot[red,mark=square,only marks,mark size=1pt] table[x index=8, y index=10] {images/fmemram3/PERF_OUTPUT_amd64_g++_o3_awk};
\addplot[red!60,mark=square*,only marks,mark size=0.6pt] table[x index=8, y index=10] {images/fmemram3_x86/PERF_OUTPUT_x86_g++_o3_awk};
\draw [gray,thin] (axis cs: 0,0) -- (axis cs: 1200000,2400000) node[pos=0.78,sloped,below] {\scriptsize $\times 2$};
\draw [gray,thin,dashed] (axis cs: 0,4299161) -- (axis cs: 1200000,6699161);
\draw [gray,->,thin,dashed,bend right,shorten <=0.1cm,shorten >=0.1cm] (axis cs: 1000000,2000000) to (axis cs: 1000000,6100000);
\draw (axis cs: 1110000, 4050000) node[gray] {\scriptsize $+\SI{4.1}{\giga\byte}$};
\end{axis}
\end{tikzpicture}
\\
    \raisebox{0.20em}{\addlegendimageintext{red,mark=square,only marks,mark size=1pt}} GCC on amd64. \quad
    \raisebox{0.20em}{\addlegendimageintext{red!60,mark=square*,only marks,mark size=0.6pt}} GCC on x86. \quad
\caption{\DerivGrind{}'s effect on the maximum resident set size of the Burgers benchmark.}
\label{fig:scalingfactor-mem}
\Description{Scatter plot with x-axis label ``maximum resident set size in MB, native execution'' and y-axis label ``max. RSS in MB, under Derivgrind''.}
\end{figure}
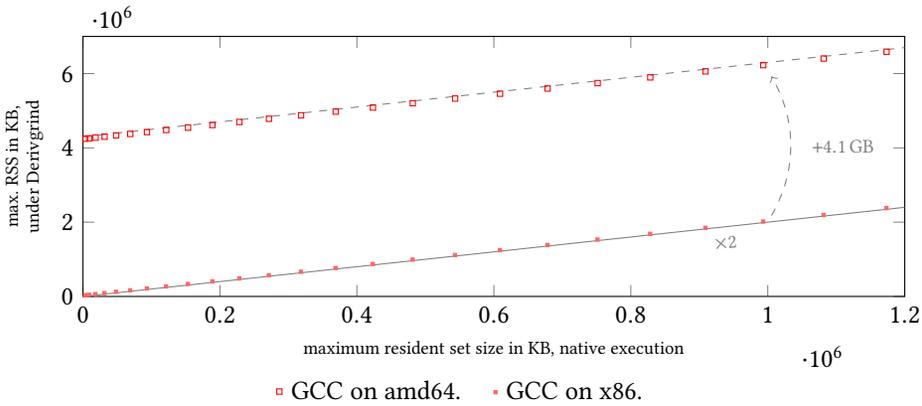

With a similar setup, \cref{fig:scalingfactor-mem} displays the effect of \DerivGrind{} on the required memory. Our memory measurements refer to the maximum resident set size (RSS) reported by the GNU \lstinline[language=bash]|time| command. We consider problem instances on an $n_x \times n_x$ grid and $n_t=4$ time steps, for $n_x=200,400,\dots,5000$, as the memory consumption hardly depends on $n_t$. As \cref{fig:scalingfactor-mem} shows, \DerivGrind{} doubles the memory consumption, in addition to a constant reservation of about \SI{4.1}{\giga\byte} on amd64 and \SI{20}{\mega\byte} on x86. On amd64, the shadow memory tool needs much more memory for its internal data structures; changing their layout, the constant allocation can be brought below \SI{0.1}{\giga\byte} with a minor run-time penalty. Compiling the program with Clang instead of GCC, and/or disabling optimizations (\lstinline[language=bash]|-O0|), had no significant effect on the required memory.

\subsection{Differentiating a Python Interpreter}\label{sec:result-python}
On many Linux systems, the default interpreter \lstinline[language=bash]|python3| for the Python programming language is CPython\footnote{\urldefCPython}. 
Core components of CPython are written in C. When the Python type \lstinline[language=python,keywordstyle={[2]\ttfamily}]|float| is used in a Python statement, CPython represents its value internally by a C \lstinline[language=c,keywordstyle=\ttfamily]|double|\footnote{\urldefCPythonPyFloatObject}. Basic arithmetic operations in Python are performed by CPython via the corresponding C operations\footnote{\urldefCPythonOperations}, and mathematical functions from the Python modules \lstinline[language=python]|math| and \lstinline[language=python]|numpy| are usually (but now always) dispatched to the C math library. 

\DerivGrind{} can thus be used to differentiate a Python script, by applying it to CPython interpreting the Python script, i.\,e.,
  \begin{equation*}
\text{\lstinline[language=bash,mathescape,basicstyle=\ttfamily]|valgrind --tool=derivgrind python3 $\phantom{a}\langle\text{\rmfamily\emph{Python script}}\rangle$ $\phantom{a}\langle\text{\rmfamily\emph{arguments for Python script}}\rangle\mathrm.$|}
\end{equation*}

In our test system based on Debian~11.6 on amd64, we obtain  \lstinline[language=bash]|python3| as a pre-built package of CPython~3.9.2 from a software repository, so the source code of CPython is not available on the system.
In order to access dot values of Python variables, we use pybind11 \cite{pybind11} to create a Python extension module \lstinline[language=python]|derivgrind|, i.\,e., a shared object  compliant with the Python/C API. At run-time, when CPython reads the Python statement \lstinline[language=python,keywordstyle=\ttfamily,breaklines=true,breakatwhitespace=true,prebreak={},postbreak={}]|import derivgrind|, it dynamically loads  \lstinline[language=bash,breaklines=true,prebreak={},postbreak={}]|derivgrind.so|. When functions of this module are used on the Python side as shown in \cref{lst:python}, CPython calls the corresponding functions of \cref{lst:clientrequest}, which were compiled into the shared object. 
In short, insertions into the source code of CPython are not necessary because CPython exposes Python variables to user-supplied C code at runtime.

\begin{lstfloat}
\caption{Usage of the Python extension module \lstinline[language=python]|derivgrind| to access the shadow memory from within a Python script, assuming the interpreter is running under Valgrind.}
\label{lst:python}
\vspace{0.5cm}
\hfill
\begin{lstlisting}[language=python,frame=single,showlines=true]
import derivgrind
x = derivgrind.set_dotvalue(4,1)
y = x*x*x
print(derivgrind.get_dotvalue(y))
\end{lstlisting}
\end{lstfloat}

We reimplementated the Burgers benchmark as a Python script, using standard Python lists instead of C arrays, \lstinline[language=python]|math.sqrt| to compute the square root, and our extension module \lstinline|derivgrind| instead of direct client request macros. We use the setup of \cref{sec:result-performance} with a $200\times 200$ grid and $200$ time steps. The computed values and derivatives match those of the \Cpp{} program. Time measurements, taken via clock reads from the Python program, result in \SI{9.5}{\second} for CPython running natively and \SI{650}{\second} for CPython running under Valgrind, corresponding to a slow-down factor of~68. 

\section{Summary}\label{sec:summary}
With the new AD tool \DerivGrind{}, we have demonstrated a methodology to augment compiled programs with forward-mode AD logic, independent of their source code. \DerivGrind{} handles x86 and amd64 machine code through the VEX intermediate representation of the Valgrind framework. Every temporary, register and memory byte is shadowed to keep track of the respective dot values. For statements with floating-point expressions or copy operations, \DerivGrind{} updates the shadows according to the respective analytical rules of differentiation.

Real arithmetic can also be  performed by integer  or logic operations, in manifold ways. Our current lack of a systematic approach to detect all the real arithmetic ``hidden'' in a portion of machine code would be a fundamental obstacle if we sought a truly universal AD tool, that could even handle hand-written assembly code of a determined counterexample-maker. However, we take a more practical perspective.  To this end, we have set up an extensive test suite, which checks various simple programs produced by the GCC and Clang compilers, as well as the precompiled CPython interpreter as it runs various Python scripts. The results indicate that unless explicitly instructed otherwise by the source code, actual compilers only very rarely realize unsupported bit-tricks.

In order to identify the input and output variables of the differentiation task, and to set or get the respective dot values, generally the parts of the client's source code containing their definitions must be accessible, in one of the following two ways. For the monitor commands interface, these parts of the source must be compiled with debugging symbols and optimization turned off. For the client request interface, they must be augmented by calls to C functions, and recompiled. If the client program exposes the variables of interest via a suitable API, this offers another way to access their dot values without any modifications of the client's source code. We used this approach to demonstrate AD for Python programs, by applying \DerivGrind{} to the Python interpreter and injecting additional C code via Python extension modules.

Time measurements on various client executables found that \DerivGrind{} scales their run-time by a factor between~30 and~75, in addition to a start-up time of a few seconds. While this is considerably slower than existing AD tools tuned for high performance, \DerivGrind{} is applicable for a much wider range of software, with less integration efforts. The real arithmetic between input and output variables is provided to \DerivGrind{} as machine code only, so it does not matter whether it has been compiled from a variety of programming languages, uses pre-compiled libraries, or comes in the shape of an interpreter running a script. Seeding inputs and retrieving output derivatives can be as easy as stopping the client in the debugger and issuing monitor commands there.


\begin{acks}
Max Aehle gratefully acknowledges funding from the research training group SIVERT by the German federal state of Rhineland-Palatinate.

We are grateful to the authors of Valgrind for creating such a highly versatile framework, and to Karl Cronburg for sharing his shadow memory library\footnote{\urldefShadowMemory} that was very helpful at an early development stage of Derivgrind.
\end{acks}

\bibliographystyle{ACM-Reference-Format}
\bibliography{lib}

\end{document}